\documentclass[twocolumn,english,prl,superscriptaddress,showpacs]{revtex4}
\usepackage[T1]{fontenc}
\usepackage[latin9]{inputenc}
\usepackage{amsmath}
\usepackage{amssymb}
\usepackage{graphicx}
\usepackage{esint}

\makeatletter
\@ifundefined{textcolor}{}
{%
 \definecolor{BLACK}{gray}{0}
 \definecolor{WHITE}{gray}{1}
 \definecolor{RED}{rgb}{1,0,0}
 \definecolor{GREEN}{rgb}{0,1,0}
 \definecolor{BLUE}{rgb}{0,0,1}
 \definecolor{CYAN}{cmyk}{1,0,0,0}
 \definecolor{MAGENTA}{cmyk}{0,1,0,0}
 \definecolor{YELLOW}{cmyk}{0,0,1,0}
 }

\usepackage{epstopdf}

\def\be{\begin{equation}}
\def\ee{\end{equation}}

\makeatother

\usepackage{babel}
\begin{document}

\title{Unified Theoretical Framework for Polycrystalline Pattern Evolution}

\author{Ari Adland}

\affiliation{Physics Department and Center for Interdisciplinary Research on Complex
Systems, Northeastern University, Boston, MA 02115}

\author{Yechuan Xu}

 \affiliation{Physics Department and Center for Interdisciplinary Research on Complex
Systems, Northeastern University, Boston, MA 02115}

\author{Alain Karma}

\affiliation{Physics Department and Center for Interdisciplinary Research on Complex
Systems, Northeastern University, Boston, MA 02115}

\date{\today}
\begin{abstract}

The rate of curvature-driven grain growth in polycrystalline materials is well-known to be limited by interface dissipation.  We show analytically and by simulations that, for systems forming modulated phases or non-equilibrium patterns with crystal ordering, growth is limited by bulk dissipation associated with lattice translation, which dramatically slows down grain coarsening. We also show that bulk dissipation is reduced by thermal noise so that those systems exhibit faster coarsening behavior dominated by interface dissipation for high Peierls barrier and high noise. Those results provide a unified theoretical framework for understanding and modeling polycrystalline pattern evolution in diverse systems over a broad range of length and time scales.  

 \end{abstract}


\maketitle

Polycrystalline patterns are observed in very diverse systems including crystalline solids \cite{SuttonBallufi1995},
colloidal systems \cite{Yoshizawaetal2011,Gokhaleaetal2012}, various spatially modulated phases of macromolecular systems such as diblock copolymers
\cite{Harrisonetal2004,Jietal2011}, and non-equilibrium (NE) dissipative structures  \cite{CrossGreenside}. 
When grain boundaries (GBs) between domains of different crystal orientation are mobile, those patterns generally coarsen in time to reduce GB length or area by elimination of smaller grains. This coarsening behavior has been extensively studied because of its practical importance for engineering polycrystalline materials \cite{HolmFoiles2010} and its fundamental relevance for our general understanding of nonequilibrium ordering phenomena.   

The ordering dynamics of modulated phases and NE patterns has been extensively studied theoretically  \cite{SwiftHohenberg1977,Elderetal1992,CrossMeiron1995,Houetal1997,BoyerVinals2001,BoyerVinals2002,Vegaetal2005,Gomezetal2007,OhnogiShiwa2011}
 in the framework of model equations of the form 
 \begin{equation}
p\partial_t^2\psi+\alpha\partial_t\psi=-(-\nabla^2)^n\frac{\delta \cal F}{\delta \psi}+\eta,~~~~(p,n ~=0~{\rm or}~1),\label{PFCeq}
\end{equation}
where $\psi$ is an order parameter appropriate to each system that can be globally conserved ($n=1$) or non-conserved ($n=0$),  $\eta$ is a noise uncorrelated in space and time with a variance determined by the fluctuation-dissipation relation $\langle \eta(\vec r,t)\eta(\vec r',t)\rangle=2\alpha T(-\nabla^2)^n \delta(\vec r-\vec r')\delta(t-t')$, and  $\cal F$ is a Lyapounov functional with a minimum in a lattice ordered state. Eq. (\ref{PFCeq}) has also been proposed as a theoretical framework$-$ the phase-field-crystal (PFC) model$-$ to study polycrystalline materials on diffusive time scales with $\psi$ representing the crystal density field \cite{Elderetal2002, Stefanovicetal2006}.
While Eq. (\ref{PFCeq}) has been traditionally studied for purely relaxational ($p=0$) dynamics \cite{SwiftHohenberg1977,Elderetal1992,CrossMeiron1995,Houetal1997,BoyerVinals2001,BoyerVinals2002,Vegaetal2005,Gomezetal2007,OhnogiShiwa2011,Elderetal2002}, propagative ($p=1$) wave-like dynamics  has also been introduced in the PFC framework to mimic phonon-mediated relaxation of the strain field \cite{Stefanovicetal2006}.  

Extensive computational studies of Eq. (\ref{PFCeq}) have shown that the characteristic domain or grain size for both roll patterns \cite{Elderetal1992,CrossMeiron1995,Houetal1997,BoyerVinals2001} and hexagonal lattices \cite{BoyerVinals2002,Vegaetal2005,Gomezetal2007,OhnogiShiwa2011}  grows $\sim t^q$. The exponent $q$ is typically much smaller than the $q=1/2$ value expected for ``normal grain growth'' in polycrystalline materials \cite{normalgg}, and depends on parameters and noise strength \cite{Gomezetal2007,OhnogiShiwa2011}. While there have been theoretical attempts to explain those exponents for roll patterns \cite{Elderetal1992,CrossMeiron1995,Houetal1997,BoyerVinals2001}, the origin of this sluggish (low $q$) coarsening kinetics is still poorly understood, especially for lattices that relate to polycrystalline materials.
  
\begin{figure}[ht]
\noindent \centering{} \includegraphics[width=7.cm]{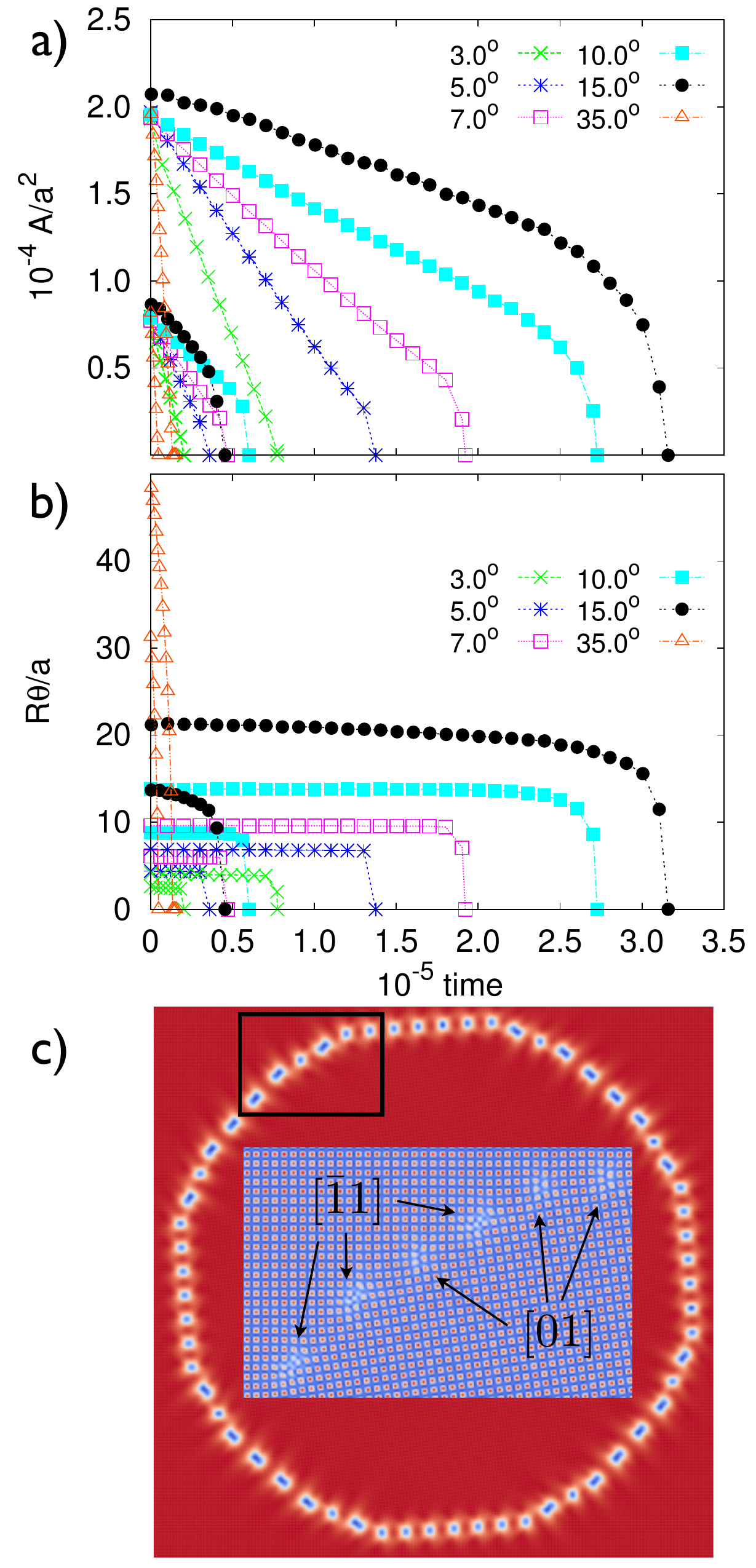} 
\caption{Results of isolated grain shrinkage simulations for $\epsilon=0.12$ and $T=0$ showing normalized grain area (a) and $R\theta/a$ (b) versus time for different initial grain areas and misorientations indicated in degrees. Grain rotation is present (absent) for misorientation less (larger) than $\sim15^o$. (c) snapshot  of defect field $\phi$ (see text) for the largest area $7^o$ simulation showing the GB structure, with $\langle 11\rangle$ ($\langle 10\rangle$) dislocation cores appearing as elongated (circular) blue patches, and the corresponding $\psi$ field inside the square region (inset).
\label{fig1}}
\end{figure}

In this letter, we show that the sluggish ordering dynamics of crystal lattices results from the subtle effect of ``bulk'' dissipation linked to lattice translation, absent in crystalline solids, but present in Eq. (\ref{PFCeq}) that lacks Galilean invariance. Internal stresses generated by GB motion are known to deform grains and even rotate them \cite{Upmanyuetal2006,CahnTaylor2004,WuVoorhees2012,TrauttMishin2012} during growth. Our main finding is that the associated dissipation can be, surprisingly, the dominant rate limiting mechanism of grain growth for a broad class of systems described by Eq. (\ref{PFCeq}). This effect is dominant in the limit where Peierls-Nabarro (PN) barriers to dislocation motion are small and curvature-driven growth does not require thermal activation. This is the physically relevant limit for lattice patterns with a wavelength much larger than the atomic scale, which coarsen near zero $T$. Interestingly, in the opposite limit of high PN barrier and high noise, bulk dissipation plays a subdominant role because noise promotes modes of GB motion that reduce grain deformation.
We summarize here our main results and will give a longer exposition elsewhere.
 
To highlight the effect of bulk dissipation, we first treat analytically the dynamics of a
circular grain of radius $R(t)$ and misorientation $\theta(t)$ with respect to its surrounding single crystal matrix. A theory for this problem has been developed by Cahn and Taylor \cite{CahnTaylor2004} for solids, and basic geometrical aspects of this theory have been validated by MD \cite{TrauttMishin2012} and PFC \cite{WuVoorhees2012} simulations.
For small initial misorientation $\theta(0)$, 
grain rotation is geometrically necessary under the assumption that the number of dislocation along the GB is conserved. Since there are $n_d=2\pi R\theta/b$ dislocations of Burgers vector magnitude $b$, this conservation condition implies that $R(\theta)\theta(t)=R(0)\theta(0)$. This rotation can also be interpreted as a consequence of the geometrical coupling between GB motion and a shear stress \cite{CahnTaylor2004,Cahnetal2006,Karmaetal2012}.  
To treat dissipation, we focus for simplicity on non-conversed dynamics ($n=0$) but the results hold generally for any $n$ or $p$. For $n=0$ and $T=0$, Eq. \ref{PFCeq} implies
\begin{equation}
\dot {\cal F}+\dot E_k  = - \int d\vec r~ \alpha(\partial_t\psi)^2,\label{Ebalance}
\end{equation}
where the dot denotes total derivative with respect to time and $E_k\equiv \int d\vec r~  (\partial_t\psi)^2/2$ is the kinetic energy of the $\psi$ field. $\dot {\cal F}$ represents the rate of change of the total GB energy of the circular grain, and thus $\dot {\cal F}=d[2\pi R(t)\gamma(\theta(t))]/dt$, where $\gamma(\theta)$ is the energy per unit length of GB. We write the total energy dissipation rate as the sum of interface and bulk contributions by splitting the integral on the right-hand-side (r.h.s.) of Eq. (\ref{Ebalance}) as the sum $\int=\int_I+\int_B$, where $\int_I$ is over a thin annulus comprising the GB of thickness proportional to the dislocation core radius, and $\int_B$ is over the bulk area of the inner grain. Dislocations move radially inward by pure glide at a velocity $\dot R$, and hence $|\partial_t\psi| \sim a^{-1}\dot R$ over an area $a^2$ around each dislocation, where $a$ is the lattice spacing. Therefore, the total interface dissipation $\int_Id\vec r\alpha(\partial_t\psi)^2 = n_da^2\alpha(a^{-1}\dot R)^2/m =  \alpha2\pi R\dot R^2\theta/(mb)$, where $m$ is an $O(1)$ dimensionless prefactor. This yields the expression for the mobility $M(\theta)=mb/(\alpha \theta)$ for $\theta\ll 1$ in agreement with previous studies \cite{CahnTaylor2004,Karmaetal2012}.

 \begin{figure}[ht]
\noindent  \centering{} \includegraphics[width=7cm]{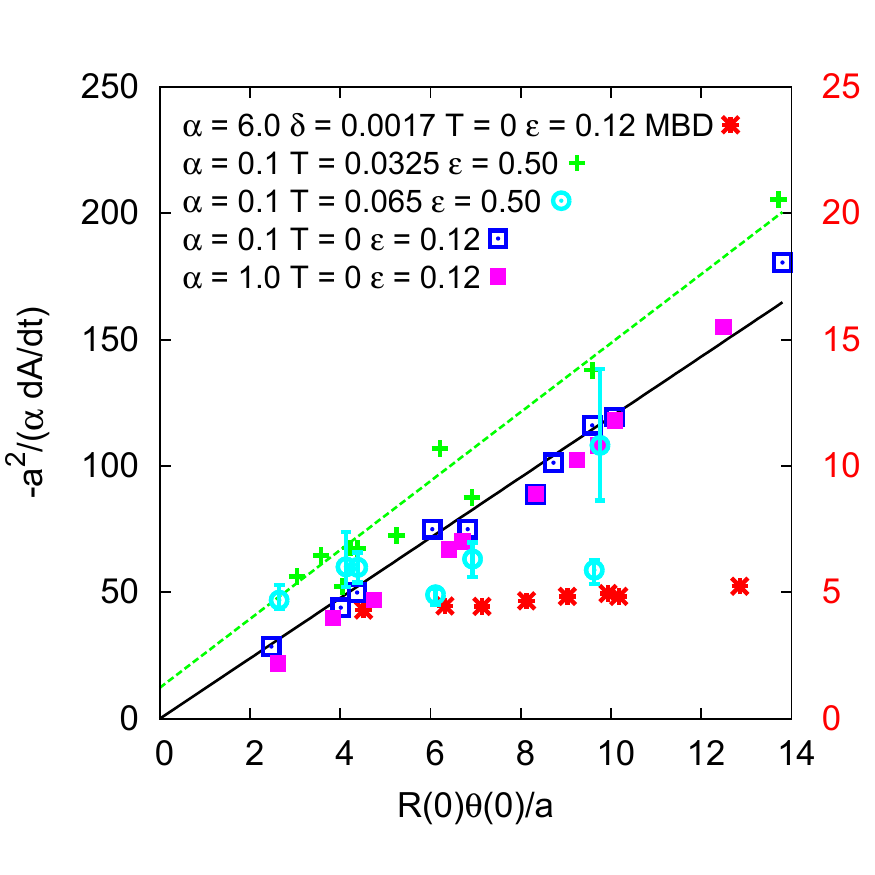} 
\caption{Comparison of theory (Eq. \ref{Adot} for the octagonal grain with two dislocation types) shown as a black solid (green dashed) line for $\epsilon=0.12$ ($\epsilon=0.5$), and simulations of Eq. (\ref{PFCeq}) (symbols). Red star symbols are for the modification of Eq. (\ref{PFCeq}) with minimized bulk dissipation (MBD), which causes the area decrease rate to become independent of grain size and misorientation for small misorientation.
  \label{fig2}}
\end{figure}

Next, to compute the bulk dissipation rate $\int_Bd\vec r~\alpha(\partial_t\psi)^2$, we compute the dissipation rate per unit area of a crystal field in uniform translation at velocity $\vec v$ and then integrate the result over the entire grain area. In a region in uniform translation, $\psi(\vec r,t)\approx \psi_0(\vec r-\vec v t)$, where $\psi_0(\vec r)$ is the equilibrium $\psi$-field that minimizes ${\cal F}$, and hence $\partial_t\psi= -\vec v\cdot\vec\nabla\psi_0$. The dissipation rate per area of crystal can therefore be written in the form $ \alpha K v^2/A_{u.c.}$ where  
$K \equiv  \int_{u.c.} d\vec r~(\hat v\cdot\vec\nabla\psi_0)^2$
is computed over the area $A_{u.c.}$ of one unit cell (u.c.). For a square lattice,  $A_{u.c.}=a^2$ and $K$ is independent of the direction of $\hat v$ relative to the crystal axes, and reduces to $K=\int_0^a\int_0^a dx'dy'(\partial_{x'}\psi_0)^2$ where $x'$ and $y'$ are the principal crystal axes. Since $v=r\dot \theta$ in each small region of a large rotating grain, where $r$ is the radial polar coordinates, 
the total bulk dissipation rate is obtained by integrating $ \alpha K v^2/A_{u.c.}$ over the grain area: $\int_0^R ~2\pi r dr \alpha K (r\dot\theta)^2/a^2= \alpha \pi K R^4 \dot\theta^2/(2a^2)=2\alpha E_k$.
Combining the above expressions for interface and bulk dissipation, Eq. (\ref{Ebalance}) becomes
\begin{equation}
\dot R\gamma+R \gamma_\theta\dot \theta=-\alpha R\dot R^2\theta/(mb)-\alpha KR^4\dot\theta^2/(4a^2),\label{dynamics}
\end{equation}
where we have neglected $\dot E_k$ which can be shown to give a negligible higher order contribution 
(so that the $p=0$ and $p=1$ dynamics in Eq. (\ref{PFCeq}) yield the same grain rotation dynamics).
Finally, approximating the GB energy with a Read-Shockley law $\gamma(\theta)=E_0\theta(A_c-\ln\theta)$ valid for small $\theta$, and using the condition $R(t)\theta(t)=R(0)\theta(0)$ that $n_d$ is conserved, Eq. (\ref{dynamics}) yields a single dynamical equation for $R(t)$ that can be analytically solved. Its solution predicts that the grain area ($A=\pi R^2$) decreases linearly in time with a rate $\dot A= dA/dt$ given by
\begin{equation}
-a^2/(\alpha\dot A)=s_1a^2/(mbE_0)+ s_2KR(0)\theta(0)/E_0,\label{Adot}
\end{equation}
where $s_1=1/(2\pi)$ and $s_2=1/(8\pi)$, and the first and second terms on the r.h.s. correspond to interface and bulk dissipation, respectively. Since $b\sim a$ and $m$ and $K$ are constants of order unity, Eq. (\ref{Adot}) predicts that the ratio of bulk to interface dissipation is $\sim R(0)/a\gg 1$, implying that $\dot A$ is entirely dominated by bulk dissipation, which holds for any lattice structure.   

\begin{figure}[ht]
\noindent \centering{} \includegraphics[width=7cm]{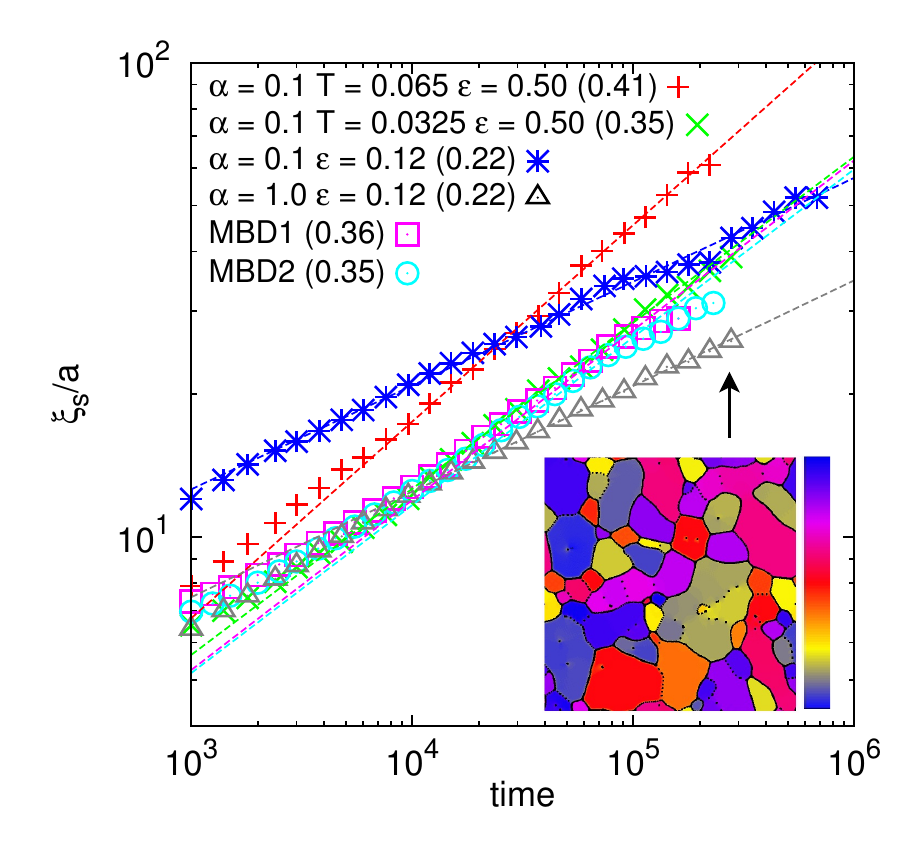} 
%\\
\caption{Plot of ordering legnth $\xi_S$ ($\sim$ mean grain size) vs. time from simulations of Eq. \ref{PFCeq} (parameters in legends) and minimized bulk dissipation with $\epsilon=0.12$, $\alpha=8$, and $\delta=3.1~10^{-3}$ ($\delta=6.3\times 10^{-3}$) for MBD1 (MBD2); $T=0$ when not given in legend. Numbers in parentheses are growth exponents from straight-line fits at large times (dashed lines). Inset: orientation field computed using a wavelet transform of $\psi$ with a color bar ranging from $-45^o$ to $45^o$ \cite{suppinfo}.\label{fig3}}
\end{figure}
 
To test this prediction, we simulated Eq. (\ref{PFCeq}) with $p=1$ and $n=0$ and ${\cal F}=\int d\vec r \,\omega$ where \cite{LifPet97,Wuetal2010} 
\begin{equation}
\omega=\psi \left[-\epsilon+(\nabla^2+1)^2(\nabla^2+2)^2\right]\psi/2+\psi^4/4-\mu\psi,\label{grandpot}
\end{equation} 
has a stable square-lattice ground state that coexists with a metastable $\psi=0$ liquid-like state for a range of $\mu<\mu_E$. This model was recently shown to produce GBs with a similar dislocation content as $[001]$ tilt GBs in molecular dynamics (MD) simulations of face-centered-cubic bi-crystals \cite{Trauttetal2012}. 
In the analogy with a crystal-liquid system, ${\cal F}$ represents the grand potential that is equal in both phases for an equilibrium value $\mu=\mu_E$ that depends on $\epsilon$
\cite{Wuetal2010}.  In the simulations reported here $\mu \approx 0.9\mu_E$, corresponding to $\mu=-0.9$ of $\epsilon=0.12$ and $\mu=-1.67$ for $\epsilon=0.5$. Since the height of the PN barrier scales $\sim \exp(-c/\epsilon^{1/2})$ where $c$ is some constant (see, e.g. \cite{BoyerVinals2001}), the height decreases rapidly with decreasing $\epsilon$. For $\epsilon=0.12$, this height is very small so that curvature-driven grain growth occurs at $T=0$ even for $R\sim 100a$. In contrast, for $\epsilon=0.5$ the barrier is large and even small grains are pinned. For $\epsilon=0.5$, we carry out simulations at finite $T$ and report results for two $T$ values to highlight noise effects. Eq. (\ref{PFCeq}) is simulated using a pseudo-spectral method described in \cite{Adlandetal2013} with more details in \cite{suppinfo}.

Fig. \ref{fig1} shows plots of grain area and $R\theta/a$ versus time together with a snapshot that highlights the structure of the GB, consisting of dislocations with Burgers vectors described by the Miller indices $\langle 11 \rangle$ and  $\langle 10 \rangle$. Accordingly, the grain is approximately shaped as an octagon with 4 facets made up of $[11]$, $[\bar 11]$, $[1\bar 1]$, and $[\bar 1\bar 1]$ dislocations, and 4 others with $[10]$, $[01]$, $[\bar 10]$, and $[0\bar 1]$ dislocations, respectively. Fig. \ref{fig1}a shows that $\dot A$ is constant and depends on both initial grain size and misorientation, and that $R\theta$ is constant. Hence, the number of dislocations is conserved until dislocations become too closely spaced and annihilate during the final stage of grain collapse. In Fig. 2, we compare quantitatively $\dot A$ values extracted from linear fits of $A$ vs. time plots in the long linear regime before grain collapse to the predictions of Eq. (\ref{Adot}) extended to an octagonal grain with two dislocation types: $E_0=(E_0^{11}+E_0^{10})/2$ and $(mb)^{-1}=[(m_{10}b_{10})^{-1}+(m_{11}b_{11})^{-1}]/2$ together with the shape constants $s_1=1/(4\sqrt{2})$ and $s_2=0.0368$ related to the perimeter and area of an octagon \cite{suppinfo}.
This comparison shows an excellent quantitative agreement for $\epsilon=0.12$ and $T=0$ 
 for two different $\alpha$ values, confirming that inertia ($p=1$) is unimportant.  
The comparison for $\epsilon=0.5$ with large PN barrier shows that bulk dissipation is still dominant for intermediate $T$ ($T=0.0325$). The slope of the curve  predicted by Eq. (\ref{Adot}) fits well the simulations results, but the curve has a finite intercept at the origin corresponding to a finite contribution of interface dissipation. This contribution is negligible for $\epsilon=0.12$ where the intercept merges with the origin. For $\epsilon=0.5$ and larger $T$ ($T=0.065$), bulk dissipation is reduced and the simulation data is no longer fitted by the theory. Analysis of dislocation dynamics in the simulations shows that this reduction results from thermally activated dislocation reactions. Those reactions reduce the number of dislocations , thereby allowing the grain to shrink with less rotation and reducing the contribution of bulk dissipation relative to interface dissipation. Reduction of grain rotation by dislocation reaction is also observed in MD simulations of grain rotation \cite{TrauttMishin2012}.

To investigate the importance of bulk dissipation in grain growth, we simulated Eq. (\ref{PFCeq}) with $p=1$ and $n=0$ in large systems of $512 \times 512$ unit cells with periodic boundary conditions. We also simulated a modified version of  Eq. (\ref{PFCeq}) that minimizes bulk dissipation (MBD) by the substitution $\alpha\rightarrow \alpha h(\phi)$ where $\phi(\vec r)=C\int d\vec r' \exp(-|\vec r-\vec r'|^2/2\zeta^2) |\nabla \psi(\vec r')|^2$ is a smoothly spatially 
varying crystalline order parameter (illustrated in Fig. \ref{fig1}b for a rotating grain), with $\zeta=a/2$ and the normalization constant $C$ chosen such that $\phi=1$ in ordered regions of the lattice and $\phi<1$ in disordered regions (dislocation cores, GBs, etc). The function $h(\phi)$ given in \cite{suppinfo} has limits $h(1)= \delta$  and $h(\phi)\sim O(1)$ in ordered and disordered regions, respectively, thereby allowing us to minimize the ratio of bulk to interface dissipation by choosing $\delta \ll 1$. Theory for MBD is identical to  Eq. (\ref{Adot}) with the second term on the r.h.s. multiplied by $\delta$. Simulation results in Fig. \ref{fig2} (red stars) confirm that bulk dissipation is negligible for $\delta\ll 1$.

We characterized the coarsening behavior by measuring the ordering scale $\xi_S$ using the half width $\delta k$ at half peak of the structure factor $S(k,t)=\langle \psi(\vec k,t)\psi(-\vec k,t) \rangle$ (fitted to a squared Lorentzian), where the angular brackets denote averages over all orientations of $\vec k$ in the same simulation \cite{CrossMeiron1995,suppinfo}. 
Plots of $\xi_S\equiv \delta k^{-1}$ vs. time in Fig. \ref{fig3} for simulations of Eq. (\ref{PFCeq}) with $\epsilon=0.12$ and $T=0$ yield a small coarsening exponent $q\approx {0.22}$ as in previous studies of hexagonal lattices \cite{BoyerVinals2002,Vegaetal2005,Gomezetal2007,OhnogiShiwa2011}. However, MBD simulations yield a substantially larger exponent $q\approx 0.35$, thereby demonstrating that bulk dissipation associated with lattice translation has a significant influence on the coarsening kinetics even though grain rotation is more constrained in a polycrystal. This conclusion is corroborated by an analysis of the simulations, which shows that grain rotation is more common with MBD and by computing the ratio of interface and total dissipation. This ratio decreases to a small value with constant $\alpha$ but remains approximately unity with MBD \cite{suppinfo}. Finally, simulations for $\epsilon=0.5$ show that the growth exponent increases markedly with $T$, as expected from the reduction of bulk dissipation by dislocation reactions. More important here than the precise values of the exponents, which are not exactly constant in time, is the demonstration of the key role of bulk dissipation in the ordering dynamics. Aspects of the present results, in particular the theoretical prediction of the striking dependence of grain area shrinkage rate on initial grain size and misorientation (Eq. \ref{Adot}), should be testable experimentally in a wide range of lattice forming systems.

This work was supported by US DOE grants DE-FG02-07ER46400.

%\bibliographystyle{aip}
%\bibliography{literat} 

\begin{thebibliography}{99}     

\bibitem{SuttonBallufi1995}
A. P. Sutton and R. W. Balluffi, {\it Interfaces in Crystalline
Materials} (Clarendon Press, Oxford, 1995).

%Exclusion of Impurity Particles during Grain Growth in Charged Colloidal Crystals. 13420Ð13427.
\bibitem{Yoshizawaetal2011} K. Yoshizawa, T. Okuzono,T. Koga, T. Taniji,and J. Yamanaka, Langmuir {\bf 27}, 13420 (2011).

%Directional grain growth from anisotropic kinetic roughening of grain boundaries in sheared colloidal crystals
\bibitem{Gokhaleaetal2012}
S. Gokhalea, K. Hima Nagamanasab, V. Santhoshc, A. K. Sooda, and Rajesh Ganapathyc, Proc. Natl. Acad. Sci. {\bf 109}, 20314 (2012). 

%Pattern coarsening in a 2D hexagonal system. 800Ð806.
\bibitem{Harrisonetal2004}
C. Harrison, D. E. Angelescu, M. Trawick, Z. Cheng, D. A. Huse, P. M. Chaikin, D. A. Vega, J. M. Sebastian, R. A. Register and D. H. Adamson,
Europhys. Lett., {\bf 67} 800 (2004).

%Domain Orientation and Grain Coarsening in Cylinder-Forming Poly(styrene-b-methyl methacrylate) Films. 4291Ð4300
\bibitem{Jietal2011}
S. Ji, C.C. Liu, W. Liao, A. L. Fenske, G. S. W. Craig, P. F. Nealey, Macromolecules {\bf  44} 4291 (2011).

\bibitem{CrossGreenside}  M. C. Cross and H. Greenside, {\it Pattern Formation and Dynamics in Nonequilibrium Systems} (Cambridge University Press, 2009).

%How grain growth stops: A Mechanism for Grain-Growth Stagnation in Pure Materials
\bibitem{HolmFoiles2010} E. A. Holm and S. M. Foiles, Science {\bf 328}, 1138 (2010).

\bibitem{SwiftHohenberg1977} J. Swift and P. C. Hohenberg, Phys. Rev. A{\bf  15}, 319 (1977).

%Ordering Dynamics in the Two-Dimensional Stochastic Swift-Hohenberg Equation. 3024-3027
\bibitem{Elderetal1992}
K. R. Elder, J. Vinals, and M. Grant, Phys. Rev. Lett. {\bf 68}, 3024 (1992).

% Domain Coarsening in Systems Far from Equilibrium
\bibitem{CrossMeiron1995} M. C. Cross and D. I. Meiron, Phys. Rev. Lett. {\bf 75}, 2152 (1995).

%Dynamical scaling behavior of the Swift-Hohenberg equation following a quench to the modulated state. 219-226.
\bibitem{Houetal1997}
Q. Hou, S. Sasa, and N. Goldenfeld, Physica A {\bf 239}, 219-226 (1997).

% Domain coarsening of stripe patterns close to onset
\bibitem{BoyerVinals2001} D. Boyer and J. Vinals, Phys. Rev. E {\bf 64} 050101(R) (2001).

%Weakly Nonlinear Theory of Grain Boundary Motion in Patterns with Crystalline Symmetry
\bibitem{BoyerVinals2002} 
D. Boyer and J. Vinals, Phys. Rev. Lett. {\bf 89}, 055501 (2002).

%Ordering mechanisms in two-dimensional sphere-forming block copolymers
\bibitem{Vegaetal2005} D. A. Vega, C. K. Harrison, D. E. Angelscu, M. L. Trawick, D. A. Huse, P. M. Chaikin, R. A. Register
Phys. Rev. E {\bf 71}, 061803 (2005).

%Effect of thermal fluctuations on the coarsening dynamics of 2D hexagonal system. 648Ð654.
\bibitem{Gomezetal2007}
L. R. Gomez, E. M. Valles, and D. A. Vega, Physica A {\bf 386}, 648 (2007).

% Effect of noise on ordering of hexagonal grains in a phase-field-crystal model
\bibitem{OhnogiShiwa2011} H. Ohnogi and Y. Shiwa, Phys. Rev. E {\bf 84}, 051603 (2011).

\bibitem{Elderetal2002} K. R. Elder, M. Katakowski, M. Haataja, and M. Grant, Phys Rev Lett {\bf 88}, 245701 (2002); K. R. Elder and M. Grant, Phys Rev E {\bf 70}, 051605 (2004); J. Berry, M. Grant and K. R. Elder, Phys Rev E {\bf 73}, 031609 (2006).

%Phase-Field Crystals with Elastic Interactions
\bibitem{Stefanovicetal2006} P. Stefanovic, M. Haataja, and N. Provatas, Phys Rev Lett {\bf 96}, 225504 (2006).

\bibitem{normalgg} The ideal grain growth $1/2$ exponent follows dimensionally from the proportionality relation between GB velocity and curvature and the assumption of a self-similar coarsening behavior (see, e.g. \cite{MullinsVinals1989}).

% Self-similarity and growth kinetics driven by surface free energy reduction.
\bibitem{MullinsVinals1989} W. W. Mullins and J. Vinals. Acta Metall. {\bf 37}, 991 (1989).

% Simultaneous grain boundary migration and grain rotation. 1707Ð1719.
\bibitem{Upmanyuetal2006}
M. Upmanyu, D.J. Srolovitz, A.E. Lobkovsky, J.A. Warren, W.C. Carter, Acta Mater. {\bf 54} 1707 (2006).

\bibitem{CahnTaylor2004} J. W. Cahn and J. E. Taylor, Acta Mater. {\bf 52}, 4887 (2004).

%Phase Field Crystal Simulations of Nanocrystalline Grain Growth in Two Dimensions, 407-419
\bibitem{WuVoorhees2012}
K.-A. Wu, and P. W. Voorhees,  Acta Mater. {\bf 60}, 407 (2012).

\bibitem{TrauttMishin2012} Z. T. Traut and Y. Mishin, Acta Mater. 60, 2407 (2012).

\bibitem{Cahnetal2006} J. W. Cahn, Y. Mishin, and A. Suzuki, Acta Mater. {\bf 54}, 4953 (2006).

\bibitem{Karmaetal2012} A. Karma, Z. T. Trautt, and Y. Mishin, Phys. Rev. Lett. {\bf 109}, 095501 (2012).
   
% Theoretical Model for Faraday Waves with Multiple-Frequency Forcing
\bibitem{LifPet97} R. Lifshitz and D. M. Petrich, Phys. Rev. Lett. {\bf 79}, 1261 (1997).

\bibitem{Wuetal2010} K.-A. Wu, A. Adland, and A. Karma, Phys Rev E {\bf 81}, 061601 (2010).

\bibitem{Trauttetal2012} Z. T. Trautt, A. Adland, A. Karma, and Y. Mishin, Acta Mater. 60, 6528 (2012); Burgers vectors are defined here differently with respect to the   crystal axes of the square lattice.

% Phase-field-crystal study of grain boundary premelting and shearing in bcc iron
\bibitem{Adlandetal2013} A. Adland, A. Karma, R. Spatschek, D. Buta, and M. Asta, Phys. Rev. B{\bf 87}, 024110 (2013).

\bibitem{suppinfo} See Supplemental Material.

 \end{thebibliography}

\end{document}